# Demonstration of Vpi Reduction in Electrooptic Modulators using Modulation Instability

**David Borlaug[1], Peter T. S. DeVore[1], Ali Rostami[2], Ozdal Boyraz[2], and Bahram Jalali[1]**

[1]Department of Electrical Engineering, University of California, Los Angeles, California 90095, USA
[2]Department of Electrical Engineering, University of California, Irvine, California 92697, USA

**Abstract:** Reduction in the operating voltage of electrooptic modulators is needed in order to remove the bottleneck in the flow of data between electronics and optical interconnects. We report experimental demonstration of a 10 fold reduction in an electrooptic modulator's half-wave voltage up to 50 GHz. This was achieved by employing optical sideband-only amplification.

## 1. Introduction

Modulation Instability (MI) is a ubiquitous nonlinear process that arises in diverse contexts such as hydrodynamics, free electron lasers, and optical rogue waves [1-3]. In fiber optics, intentionally stimulating MI leads to breakup of continuous wave beams or long pulses into high repetition rate pulse trains [4], and to improvements in stability of supercontinuum light sources [5-12]. In this paper, we experimentally show how this phenomenon can dramatically boost electrooptic (EO) modulation. In the so-called modulation instability booster (MiBo), modulation sidebands stimulate MI in a third-order nonlinear optical material placed after the EO device. This results in an effective enhancement of the electrooptic response and hence a reduction in the characteristic drive voltage, $V_\pi$. It also compensates for the high-frequency roll-off and extends the modulator's operating bandwidth. This technique was recently proposed and theoretically analyzed as a mean to enable low-voltage electrooptic modulation at ultrahigh frequencies [13]. Here we report the first experimental demonstration of $V_\pi$ reduction using this method.

Improving EO modulation is critical because the trend in electronics toward low-voltage high-frequency operation is in conflict with the increase in the drive voltage of EO modulators at high frequencies. As a consequence, EO modulation has become the bottleneck in flow of data between electrical and optical domains. Unfortunately, the development of modulators capable of simultaneous low-voltage and high-frequency operation remains elusive due to the intrinsic tradeoff between low-voltage and wide bandwidth. To achieve low voltage, a long electrooptic interaction length with traveling wave electrodes is needed. But this increases the walk off between optical and microwave waveguides – a detriment that becomes more severe with increasing bandwidth.

This paper utilizes a technique whereby an electrooptic booster is inserted following the modulator. The booster functions by increasing the amplitude of modulation sidebands at the expense of the carrier. By increasing the modulation depth for a given applied voltage, this technique enables low-voltage high-frequency operation. Our solution is applicable to intensity modulated analog links and to digital links that employ return-to-zero modulation format.

## 2. Principle of Operation

Modulation instability (MI) in optics results from the interplay between Kerr nonlinearity and anomalous group velocity dispersion. The effect can be obtained using linear stability analysis performed on analytical solutions to the nonlinear Schrodinger equation:

$$\frac{\partial A(z,t)}{\partial z} = \underbrace{-\frac{\alpha}{2}A(z,t)}_{\text{linear loss}} + \underbrace{\sum_{k\geq 2}\frac{i^{k+1}}{k!}\beta_k\frac{\partial^k A(z,t)}{\partial t}}_{\text{linear dispersion}} + \underbrace{i\gamma(1-f_R)A(z,t)|A(z,t)|^2}_{\text{Kerr nonlinearity}} + \underbrace{i\gamma f_R A(z,t)\int_{-\infty}^{t}h_R(t-t')|A(z,t)|^2\,dt'}_{\text{Raman nonlinearity}} \quad (1)$$

where $A(z,t)$ is the field amplitude, $\alpha$ is the linear loss coefficient, $\beta_k$ are the dispersion coefficients, $\gamma$

is the third-order nonlinear coefficient, $f_R$ is the fraction of third-order nonlinearity contributed by the Raman effect, and $h_R$ is the Raman response function [4]. Over electrical signal bandwidths of interest (less than 1 THz) the Raman effect and high-order dispersion terms, $\beta_k$ for $k > 2$, can be ignored. The stochastic nature of MI involves rather complex dynamics, including temporally-confined MI [14], MI with near-zero $\beta_2$ [15], and MI where the sidebands grow too large [16–18]. However, the initial evolution of MI from continuous-wave radiation with sizable anomalous group velocity dispersion can be described using a simple analytical form. In this case the sideband gain per unit length is [4]:

$$g(\omega) = |\beta_2 \omega|[(4\gamma P/|\beta_2|) - \omega^2]^{1/2} \quad (2)$$

where $\omega$ is the pump-sideband frequency separation and $P$ is the pump optical power. In the case of modulation seeded MI, the pump-sideband frequency separation is simply the radio frequency (RF). The gain shape is shown in fig. 1(a). The gain first increases linearly with frequency, peaks at $\omega_{peak} = \sqrt{4\gamma P/|\beta_2|}$, and then rolls off. The gain shape can be adjusted to compensate the EO modulator's roll-off by choosing the correct combination of the above coefficients, ultimately enabling low-voltage broadband modulation, c.f. fig. 1(b). In this paper the RF frequency is instrument limited to 50 GHz. However, the large and tunable bandwidth of MI ( > 10 THz [4] ) can be extended well beyond the bandwidths presented here.

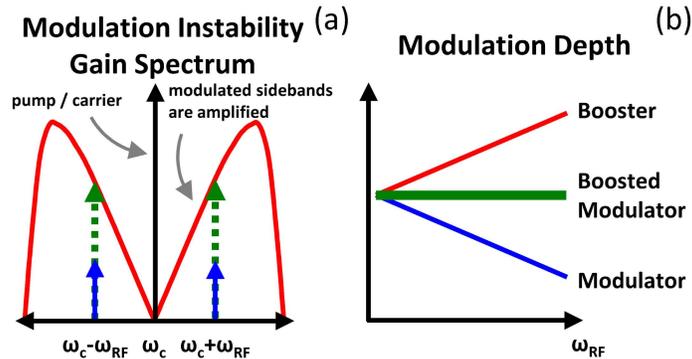

Fig. 1. (color online) (a) Principle of electrooptic modulation enhancement using modulation instability. Modulation sidebands are boosted due to MI. Note that the gain increases with RF frequency. (b) Diagram of modulation depth vs. RF frequency. A low-voltage high-frequency modulator is achieved as the booster inherently compensates the high-frequency roll-off of the modulator.

## 3. Experimental Setup

The experimental setup is shown in figure 2. A tunable laser diode at 1550 nm feeds an isolator protected EDFA, the output of which is polarization aligned with a phase modulator that is driven by an amplified 2.5 GHz band-limited white noise source for stimulated Brillouin scattering (SBS) suppression. The amplified noise source has a peak-to-peak voltage of 6.32 V and drives an EOSpace 10 GHz phase modulator, model number PM-0K1-10-PFU-PFU, having a half-wave voltage of 4.0 V at 1 GHz. The SBS suppressed carrier is polarization aligned with an EOSpace 20 GHz Mach-Zehnder modulator, model number AZ-1x2-0K1-20-PFU-SFU, having a half-wave voltage of 4.5 volts at 1 GHz. The amplitude modulator is driven by an HP 83650B synthesizer at RF frequencies from 10 to 50 GHz with 2.5 GHz granularity. The RF output power is kept constant at 1 mW. The modulated waveform is amplified by an isolator protected EDFA, variably attenuated, boosted by the highly-nonlinear fiber, and detected by an optical spectrum analyzer. At 1550 nm the HNLF is described by the following coefficients: $\alpha$ = 1 dB/km, $\beta_2$ = −10.9 ps$^2$/km, $\beta_3$ = 0.062 ps$^3$/km, and $\gamma$ = 11.6 W$^{-1}$km$^{-1}$. A circulator and various power monitors are used to precisely control the power before entering the HNLF and to monitor SBS suppression.

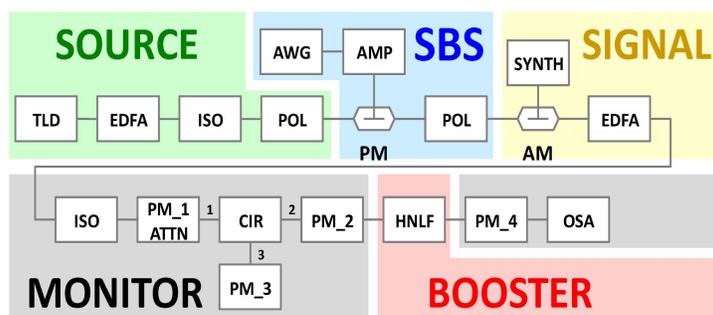

Fig. 2. (color online) Experimental setup. A continuous-wave laser source (SOURCE), acting as the carrier and pump, is phase modulated to prevent stimulated Brillouin scattering (SBS) in the BOOSTER. The RF signal is modulated onto the carrier (SIGNAL) and is boosted in the BOOSTER. The MONITOR measures the booster input power, output power, back-reflected power, and optical spectra. The MONITOR is therefore able to verify the limited impact of stimulated Brillouin scattering on the boosting process. TLD: tunable laser diode; EDFA: erbium doped fiber amplifier; ISO: isolator; POL: polarization controller; PM: phase modulator; AWG: arbitrary waveform generator; AMP: amplifier; AM: amplitude modulator; SYNTH: synthesizer; PM_[#]: power meter, numeric designator; CIR: circulator; HNLF: highly nonlinear fiber of 2 km length; OSA: optical spectrum analyzer.

## 4. Noise Analysis

Understanding the noise and linearity of MiBo is necessary when selecting to use MiBo or a combination of MiBo and compatible methods, such as RF-waveform pre-amplification, and/or unanticipated low-voltage broad-bandwidth modulator development. A few comments on noise are made in this section. Ultimately the noise performance of the MiBo approach must be measured experimentally to understand the value of the approach, which will be the focus of future work.

The MiBo architecture works by seeding MI using intensity modulation sidebands. For weak sidebands, amplified vacuum noise competes with sideband boosting and is therefore the dominant noise source in this region of operation. For the strong sideband region of operation the dominant noise sources are pump depletion, sideband intermixing, and cascaded mixing. The MiBo approach must operate in the region where the sidebands are strong enough that they beat out MI-based ASE and weak enough such that they do not deplete the carrier or undergo mixing. With regard to intermixing and cascaded mixing, it is important to note that intensity modulators such as Mach-Zehnder modulators have inherent third-order nonlinearities and second-order nonlinearities when quadrature bias errors are present. Future MiBo noise measurements must therefore take into account the contribution of noise and distortion from MI and from the modulator itself. Additionally, the carrier power and the nonlinear element length must be chosen to: (1) discourage mixing and, (2) ensure that the sidebands are preferentially amplified over MI-based ASE. In addition to MI-based noise sources, electrical noise sources such as thermal and shot noise, as well as optical noise such as phase and relative intensity noise must be considered.

Apart from the mentioned noise sources, technical considerations such as the use of EDFAs and circulators for SBS monitoring can impact noise performance. The experimental setup uses 2 km of HNLF via cascaded 1000 m, 700 m, and 300 m sections. We chose multiple sections for experimental flexibility and we protected each fiber with sacrificial patch cords to prevent facet burning from ruining the HNLF. This has resulted in a total excess loss of 6 dB beyond that of the HNLF. In other words, we're throwing away 6 dB of power for experimental flexibility and long term lab equipment care. If we were to remove these splices as would be done for a commercial product our 200 mW input power requirement (discussed in the results section) would translate to a 50 mW input power requirement. In the following section we will show $V_\pi$ improvement at input power levels as low as 100 mW, which would translate to an input power of 25 mW for a commercial fiber. The reason for utilizing the EDFA in the current setup is to overcome the additional 4 dB of loss from the circulator before entering the HNLF and the 6 dB of loss from the protective patch cords. In this paper the circulator is critical to our important SBS suppression monitoring. However, in a finely tuned commercial product the monitoring would not be necessary. Overall, our experimental demands on input power are 10 dB higher than physically required due to SBS monitoring and HNLF protection. It's worth mentioning that as of this writing the company NP Photonics advertises turnkey 80 and 125 mW ultra low noise lasers [19]. Due to the above, the MiBo approach is not EDFA reliant, and future noise projections and experiments can consider EDFA-free operation. That being said, utilizing an EDFA in our current experiments is convenient as it allows for our sacrificial patch cords to protect our costly fibers and allows for SBS monitoring.

Using the small signal gain described in equation (2) some comments can be made with regard to the relative intensity noise (RIN), and the amplified spontaneous emission (ASE) originating from laser and optical amplifiers. Since the modulation side lobes are generated from the carrier, all frequency components start with identical RIN values. RIN values can be as low as -165 dB/Hz for a standard telecommunications laser which will need to be amplified by an EDFA to obtain the required power levels. Due to the nonlinear behavior of MI gain, small-signal pump fluctuations due to RIN ($P = P_0 + \Delta P$) will result in parametric gain ($G = G_0 + \Delta G$) where $\Delta G$ is a complex function of $P$. The root-mean-square (RMS) value of the amplitude fluctuations at the signal wavelength can be calculated as [20]:

$$\sigma^2_{signal} = \langle P^2_{signal} \rangle - \langle P_{signal} \rangle^2 = \left\{ \frac{dG}{dP}\bigg|_{P=P_0} \right\}^2 \sigma^2_{pump-RIN} \tag{3}$$

where $\sigma^2_{pump-RIN}$ describes the RIN noise present at the pump frequency and it is described as $\sigma^2_{pump-RIN} = RIN \cdot (R \cdot P_0)^2 \Delta f$ where $R$ is the responsivity of the photodetector and $\Delta f$ is RF bandwidth. Figure 3 below illustrates the total RIN of the side lobes as a function of frequency for a DFB pump laser with RIN of -165 dB/Hz. The RIN degradation increases as a function of frequency with increasing MI gain. While future experiments may operate without the use of an EDFA, some comments regarding EDFA-based ASE noise are made here. These comments are especially relevant for amplified telecommunications lasers. In particular, experimental procedures starting with DFB lasers will require fiber amplifiers that unfortunately introduce broadband ASE with a power spectral density of $S_{ASE} = n_{sp} h \nu (G_{amp} - 1)$. Here $n_{sp}$ defines the spontaneous emission factor and $G_{amp}$ describes the optical gain. The presence of ASE creates amplitude fluctuations at the pump and signal wavelengths. The initial amplitude noise at the pump frequency will be $\sigma^2_{pump-ASE} = S_{ASE} (2R\sqrt{P_0})^2 \Delta f$. Such amplitude fluctuations will then be amplified by the MI process. The dotted red line in figure 3 illustrates the noise contribution of ASE from a 13 dB gain optical amplifier with 5 dB noise figure after MI boosting. The impact of ASE becomes more pronounced at higher frequencies.

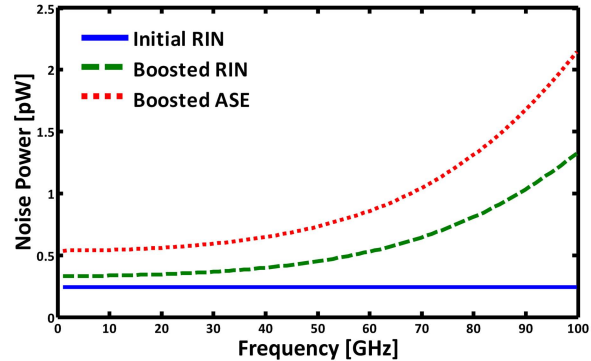

Fig. 3. (color online) The solid blue line represents the initial RIN present before MI gain as a function of RF frequency. The dashed green line represents RIN after the MI booster as a function of frequency. The dotted red line represents the boosted ASE noise that was introduced by the EDFA.

The signal-to-noise ratio (SNR) degradation is attributed here to these two noise components, RIN and ASE. Figure 4 below shows the SNR degradation for the following system: a laser with -165 dB/Hz RIN, followed by a 13 dB optical amplifier with 5 dB noise figure, followed by the HNLF described in the above sections. The overall expected SNR degradation due to these noise sources is bounded between 1.5 dB to 6 dB over the frequency range of DC to 100 GHz ( c.f. figure 4 ). This degradation can further be reduced by more than 3dB at 100 GHz by using a narrowband fiber-based laser with high power output [19]. Such lasers will eliminate the need of optical amplification and hence get rid of the amplitude noise induced by ASE.

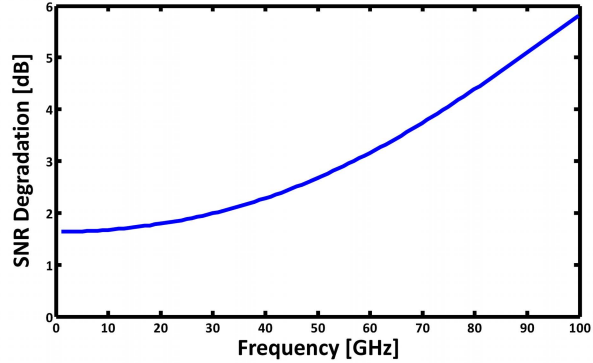

Fig. 4. (color online) Relative SNR showing degradation due to the combined effects of RIN and EDFA-based ASE. At 50 GHz the SNR is degraded by 3.5 dB.

These results indeed indicate that the ASE is the biggest contributor in SNR degradation. To further assess the affect of ASE and degenerate FWM process we use split step Fourier method to solve the nonlinear Schrodinger equation that includes all pertinent dispersion and all possible nonlinear mixing process between the signal and noise. In these simulations we assume an EDFA model with tunable gain and constant 5 dB noise figure. To verify experimental results the simulations are performed for 1 mW, 50 mW and 200 mW pump powers. As simulation results indicate in Fig 5, while gain increases in the EDFA the MI process starts with higher ASE seed at the beginning of the fiber. Through the MI process this seeded ASE is further amplified to contribute higher SNR degradation as summarized in Fig 3. We also simulate the MI system with direct laser output without EDFA and show that MiBo with low SNR degradation is possible. However, as illustrated in experimental observations and in Fig 5, degenerate four wave mixing creates higher order sidebands that may pose additional noise degradation as gain approaches 10 dB. These effects warrant further study particularly for high gain MI and broadband applications where higher harmonics may overlap with signal frequencies.

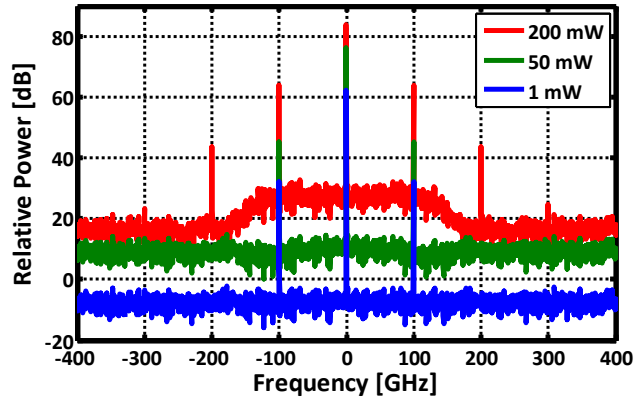

Fig. 5. (color online) Simulated effect of ASE noise in MiBo. Amplification of the pump seeds broadband ASE that is amplified by the MI process along with signal and degrades the SNR of high frequency sidebands.

The estimated noise performance indeed assumes a pump laser with -165 dB/Hz RIN. However, RIN is rarely characterized at frequencies beyond 10 GHz. At the higher frequency range pertinent to the present work, we can only rely on analytical RIN models developed for semiconductor lasers [21]:

$$RIN(f) = \frac{Af^2 + B}{(f_r^2 - f^2) + Cf^2} \qquad (4)$$

Here $A$ and $B$ are constants depending on Langevin noise sources, $f_r$ is the relaxation oscillation frequency and $C$ is a parameter that describes the damping of relaxation oscillations. Potential laser sources for future experimental setups include semiconductor DFB lasers with -165 dB/Hz RIN and commercial high-power narrow-linewidth fiber lasers [19].  Based on curve fitting of a particular commercial high-power fiber laser's spec sheet data [19], at high frequencies RIN values as low as -200 dB/Hz (well below the measurement limit) may be utilized to avoid RIN degradation.  By comparison, equation (4) estimates RIN values below -180 dB/Hz at frequencies above 40 GHz for a DFB laser with the following characteristics: a resonance peak near 3 GHz, a peak RIN value of -150 dB/Hz, and RIN below -160 dB/Hz at 10 GHz.  These results indicate that the actual RIN degradation at higher frequencies might be substantially lower than those estimated in Figure 3 and that the ASE noise source can be eliminated entirely through the use of commercial high-power narrow-linewidth fiber lasers [19].

## 5. Experimental Results

Before discussing the core MiBo results it is critical to understand role of Brillouin scattering in the experiments.  Figure 6 shows the transmitted and reflected power as a function of the launch power with respect to the HNLF.  These power levels correspond to PM_4, PM_3, and PM_2 of figure 2, respectively.  The launch power is set manually by adjusting the variable attenuator in PM_1/ATTN.  Two cases are shown: (1) the case where SBS suppression is turned off as shown using the dashed line, and (2) the case where SBS suppression is turned on as shown using a solid line.

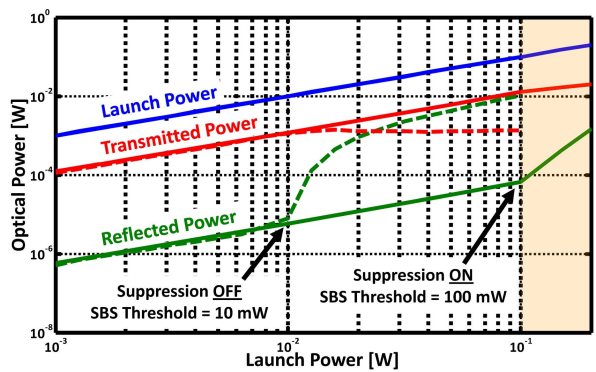

Fig. 6. (color online)  Monitoring of SBS with suppression on (solid lines) and off (dashed lines).  Note that the deviation from linear reflected power as a function of launch power occurs at 10 mW for the unsuppressed case and 100 mW for the suppressed case.  With SBS suppression on, SBS is completely suppressed up to and including 100 mW launch powers.  At 200 mW launch power 3.4 mW is reflected by SBS.

As seen in the figure, SBS suppression is successful in regions that observe a linear dependence between the reflected power and the launch power [22].  In the case of SBS suppression turned off, SBS onset begins beyond 10 mW launch power.  In the case of SBS suppression turned on, SBS begins beyond 100 mW launch power.  The approach achieves 10 dB of SBS suppression.  The reader will please note that SBS will play no role in experiments utilizing launch powers up to and including 100 mW with SBS suppression turned on.  At 200 mW launch power with SBS suppression turned on SBS causes a 3.4 mW back reflection and a 7 mW drop in transmitted power.  Suppression of SBS is critical to ensuring observed modulation enhancement is due to MI, as desired.  Past works have shown as high as 17 dB SBS suppression using similar approaches [22].  We believe our suppression capability can be improved upon by utilizing a phase modulator with an integral polarization analyzer.  Such a modulator would prohibit any un-suppressed light from making its way to the HNLF, which we believe is the cause of the reduced performance.  For the following experiments SBS suppression is turned on.

   Now that the SBS suppression capabilities are understood further measurements can proceed.  The power entering the HNLF was set to { 1, 50, 100, 150, 200 } mW and a computer was used to synchronously sweep the RF frequency and to grab OSA traces.  The experimental OSA traces are shown in figure 7.  The drawn to scale insert shows the modulation envelope of the RF sideband at 1 mW carrier power and at 200 mW carrier power. Inset spectra are carrier normalized to permit comparison of the modulation depth.  Notice the improved modulation depth with respect to the carrier and the flattening of the frequency response as the carrier power increases.  The rise in the noise floor is due to the increase in EDFA ASE noise.  The amplification of the ASE by MI gain spectrum is also evident.  Indeed the rise of the noise floor with EDFA gain and MI gain can be predicted, as illustrated in Fig 5, by numerically solving the nonlinear Schrodinger equation. Use of a high power signal laser capable of providing up to 200 mW power would eliminate the need for EDFA. The satellite peaks on top of the MI amplified ASE is believed to be due to FWM, as predicted in the previous section.

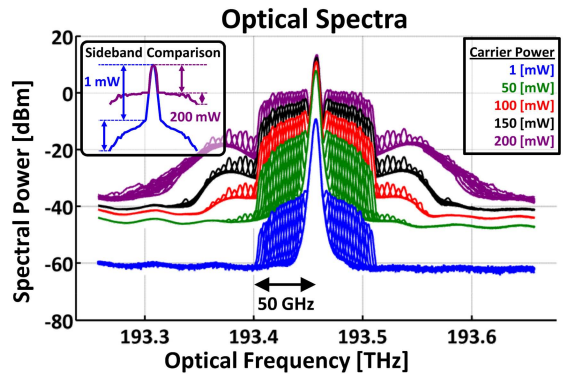

Fig. 7. (color online) Experimental OSA traces. Seventeen individual traces, one for each RF frequency, are taken for each power level and plotted in a single color. Drawn to scale inset compares carrier normalized modulation envelope of 1 mW RF sideband and 200 mW RF sideband. Evident are improved modulation depth and compensation of the modulator's frequency roll off at high modulation (sideband) frequencies.

The half-wave voltage, or $V_\pi$, was extracted from the experimental OSA traces via comparison to a numerically simulated OSA trace of a linear optical link. The simulated linear link consisted of an ideal laser connected to an ideal Mach–Zehnder modulator detected by an OSA. The effective half-wave voltage was determined by numerically varying both the simulated laser power and the simulated half-wave voltage until the error between the experimental and simulated OSA traces was minimized. Figure 7 shows the effective half-wave voltage of the boosted modulator as a function of RF frequency and carrier power.

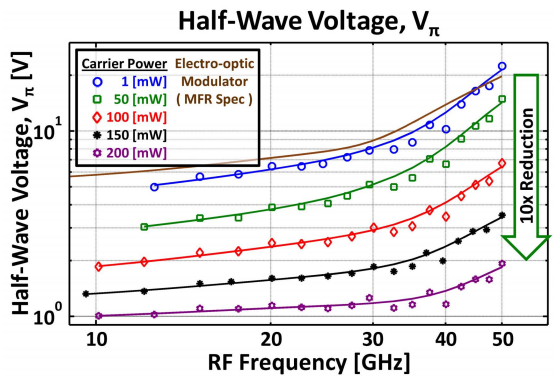

Fig. 7. (color online) Effective half-wave voltage of boosted modulator as a function of RF frequency and carrier power. Note the EO modulator's manufacturer specification. For 200 mW the half-wave voltage is reduced by more than 5-fold at 10 GHz, at 50 GHz the reduction is 10-fold.

At 1 mW carrier power (blue line) the experimentally measured effective half-wave voltage closely matches the manufacturer's specification (brown line). As the carrier power is increased there is an overall reduction in the half-wave voltage as well as improved flatness over the measurement range. Note that this trend is present in all cases of increased carrier power, including 100 mW carrier power (red line) where SBS is fully suppressed (c.f. Fig. 7). This shows that the effect is due to MI as desired. The flatness can be measured by examining the ratio of largest to smallest half-wave voltage across the 10 to 50 GHz range. For low optical power the ratio is approximately 3.3. At 200 mW carrier power this ratio is reduced to 2, evidencing the ability of MiBo to equalize the RF response of the modulator. Note that the required signal power requirements scale as the square of the voltage. For carrier power of 200 mW the half-wave voltage is reduced by more than 5-fold at 10 GHz and by 10-fold at 50 GHz to 1 V and 2 V, respectively.

## 6. Discussion

In summary, it was experimentally demonstrated that modulation instability can be exploited to enable low-voltage high-frequency electrooptic modulation. The demonstration resulted in a 10-fold reduction of the effective half-wave voltage $V_\pi$ at 50 GHz. Future work must consider the impact of modulation instability on the noise and dynamic range of the communication link.

## Acknowledgements

This work was supported by the Defense Advanced Research Projects Agency (DARPA) MTO and SSC Pacific via contract no. N66001-14-1-4001, the Office of Naval Research (ONR), contract no. N00014-12-1-0025, and Sandia National Labs through a graduate research fellowship.

## References

[1] T. B. Benjamin and J. E. Feir, J. Fluid Mech. 27, 417 (1967).
[2] D. R. Solli, C. Ropers, P. Koonath, and B. Jalali, Nature (London) 450, 1054 (2007).
[3] D. R. Solli, G. Herink, B. Jalali, and C. Ropers, "Fluctuations and correlations in modulation instability," Nature Photonics 6. 463 (2012).
[4] G. P. Agrawal, Nonlinear Fiber Optics (Academic Press, Boston, 2007).
[5] D. R. Solli, B. Jalali, and C. Ropers, "Seeded supercontinuum generation with optical parametric down-conversion," Physical Review Letters 105, 233902 (2010)
[6] D. R. Solli, C. Ropers, and B. Jalali, "Active control of rogue waves for stimulated supercontinuum generation," Physical Review Letters 101, 233902 (2008).
[7] J. M. Dudley, G. Genty, and B. J. Eggleton, Opt. Express 16, 3644 (2008).
[8] P. T. S. DeVore, D. R. Solli, C. Ropers, P. Koonath, and B. Jalali, Appl. Phys. Lett. 100, 101111 (2012).
[9] J. M. Dudley, G. Genty, and S. Coen, Rev. Mod. Phys. 78, 1135 (2006).
[10] X. Xu, C. Zhang, T. I. Yuk, K. K. Tsia, K. K. Y. Wong, "Stabilized wide-band wavelength conversion enabled by CW-triggered Supercontinuum," IEEE Photonics Technology Letters 24, 1886-1889 (2012)
[11] Qian Li, Feng Li, Kenneth K. Y. Wong, Alan Pak Tao Lau, Kevin K. Tsia, and P. K. A. Wai, "Investigating the influence of a weak continuous-wave-trigger on picosecond supercontinuum generation," Opt. Express 19, 13757-13769 (2011)
[12] K. K. Y. Cheung, C. Zhang, Y. Zhou, K. K. Y. Wong, and K. K. Tsia, "Manipulating supercontinuum generation by minute continuous wave," Opt. Lett. 36, 160-162 (2011)
[13] Peter T. S. DeVore, David Borlaug, and Bahram Jalali, " Enhancing electrooptic modulators using modulation instability," Phys. Status Solidi RRL 7, No. 8, 566–570 (2013).
[14] D. R. Solli, G. Herink, B. Jalali, and C. Ropers, Nature Photon. 6, 463 (2012).
[15] M. Droques, B. Barviau, A. Kudlinski, M. Taki, A. Boucon, T. Sylvestre, and A. Mussot, Opt. Lett. 36, 1359 (2011).
[16] N. N. Akhmediev and V. I. Korneev, Theor. Math. Phys. 69, 1089 (1986).
[17] J. M. Dudley, G. Genty, F. Dias, B. Kibler, and N. Akhmediev, Opt. Express 17, 21497 (2009).
[18] M. Taki, A. Mussot, A. Kudlinski, E. Louvergneaux, M. Kolobov, and M. Douay, Phys. Lett. A 374, 691 (2010).
[19] NP Photonics, "Rock Source: Compact Single-frequency Benchtop Fiber Laser Source," product specification sheet, http://www.npphotonics.com/images/pdfs/products/10_Rock_Laser_Source.pdf, accessed 2014-04-04.
[20] Per Kylemark, Per OlofHedekvist, et al, Journal of lightwave Technology, V 22, N 2, pp 409-416, 2004
[21] C. Carlsson, " Design and evaluation of vertical cavity surface emitting lasers for microwave applications", Ph.D. Dissertation, Photonics Laboratory, Department of Microtechnology andNanoscience (MC2), Chalmers University of Technology, Göteborg, 2003. ISSN: 91-7291-351-7
[22] J. B. Coles, et. al., "Bandwidth-efficient phase modulation techniques for Stimulated Brillouin Scattering suppression in fiber optic parametric amplifiers," Optics Express, 18(17), 18138 (2010).
[23] M.N. Islam, and O. Boyraz, IEEE J. Selected Topics in Quantum Mechanics, V8, pp-527-537, 2002.